\documentstyle [a4p,epsf,12pt] {article}
\catcode`@=11 
\def \gsim{\mathrel{\mathpalette\@versim>}}
\def \lsim{\mathrel{\mathpalette\@versim<}}
\def \@versim#1#2{\lower0.4ex\vbox{\baselineskip\z@skip\lineskip\z@skip
     \lineskiplimit\z@\ialign{$\m@th#1\hfil##\hfil$%
     \crcr#2\crcr\sim\crcr}}}
\catcode`@=12 
\def\Bp{\mathrm{B^+}}
\def\B0{\mathrm{B^0}}

\def\Bd{\mathrm{B_d^0}}
\def\Bdb{\mathrm{\bar B_d^0}}
\def\taud{\tau_{\mathrm d}}
\def\Bs{\mathrm{B_s^0}}
\def\Bsb{\mathrm{\bar B_s^0}}
\def\BB{\mathrm{B - \bar B}}

\def\ccbar{\mathrm{c\overline{c}}}
\def\bbbar{\mathrm{b\overline{b}}}

\def\uubar{\mathrm{u\overline{u}}}\def\ddbar{\mathrm{d\overline{d}}}
\def\ssbar{\mathrm{s\overline{s}}}
\def\arr{\rightarrow}
\def\Z{\mathrm Z^0}
\def\Zbb{\mathrm{Z \arr \bbbar}}
\def\Zcc{\mathrm{Z \arr \ccbar}}

\def\dmd{\Delta m_{\mathrm{d}}}
\def\dms{\Delta m_{\mathrm{s}}}

\def\Lb{\Lambda_{\mathrm{b}}}
\def\bpr{\mathrm{b \arr \ell}}
\def\bcas{\mathrm{b \arr c \arr \ell}}
\def\bcb{\mathrm{b \arr \bar{c} \arr \ell}} 
\def\btau{\mathrm{b \arr \tau \arr \ell}} 

\def\Gc{\rm GeV/$c$}
\def\etal{{\sl et al.}}
\def\nn{\alpha_{\mathrm{kin}}}
\def\bhemi{\beta_{\mathrm hemi}}
\def\btag{\beta_{\mathrm evt}}

\def\ps{ps$^{-1}$}

\begin{document}           
\begin{titlepage}
\begin{center}
  {\large   EUROPEAN LABORATORY FOR PARTICLE PHYSICS }
\end{center}
\bigskip
\begin{tabbing}
\` CERN-EP/99-085 \\
\` 22 June 1999 \\
\end{tabbing}
\vspace{1 mm}
\begin{center}{\LARGE\bf
A Study of 
\mbox{\boldmath $\Bs$} Meson Oscillation
}\end{center}
\begin{center}{\LARGE\bf
\mbox{\boldmath Using Hadronic $\Z$ Decays}
}\end{center}
\begin{center}{\LARGE\bf
Containing Leptons
}\end{center}
\vspace{10 mm}
\begin{center}{\LARGE
The OPAL Collaboration
}\end{center}
\vspace{6 mm}
\begin{abstract}
\vspace{5 mm}
A sample of $\Z$ decays 
containing b-flavoured hadrons
is tagged using leptons,
and events having precise proper time measurements
are selected.
These events are used to study $\Bs$ oscillations.
The flavour (b or $\rm \bar{b}$) at decay is determined
from the lepton charge while the flavour at production
is determined
from jet charge or the charge of a second lepton, where available.
The experiment was not able to resolve the oscillatory
behaviour, and we deduce that the $\Bs$ oscillation frequency
$\dms > 5.2$~\ps\ at the 95\% confidence level.
\parskip 0.5cm
\end{abstract}
\vspace{10mm}
\begin{center}
(Submitted to European Physical Journal C)
\end{center}
\end{titlepage}
\begin{center}{\Large        The OPAL Collaboration
}\end{center}\bigskip
\begin{center}{
G.\thinspace Abbiendi$^{  2}$,
K.\thinspace Ackerstaff$^{  8}$,
G.\thinspace Alexander$^{ 23}$,
J.\thinspace Allison$^{ 16}$,
N.\thinspace Altekamp$^{  5}$,
K.J.\thinspace Anderson$^{  9}$,
S.\thinspace Anderson$^{ 12}$,
S.\thinspace Arcelli$^{ 17}$,
S.\thinspace Asai$^{ 24}$,
S.F.\thinspace Ashby$^{  1}$
% ...
}\end{center}\bigskip
\bigskip
$^{  1}$School of Physics and Astronomy, University of Birmingham,
Birmingham B15 2TT, UK
\newline
$^{  2}$Dipartimento di Fisica dell' Universit\`a di Bologna and INFN,
I-40126 Bologna, Italy
\newline
% ...
$^{  5}$Cavendish Laboratory, Cambridge CB3 0HE, UK
\newline
% ...
$^{  8}$CERN, European Organisation for Particle Physics,
CH-1211 Geneva 23, Switzerland
\newline
$^{  9}$Enrico Fermi Institute and Department of Physics,
University of Chicago, Chicago IL 60637, USA
\newline
% ...
$^{ 12}$Indiana University, Department of Physics,
Swain Hall West 117, Bloomington IN 47405, USA
\newline
% ...
$^{ 16}$Department of Physics, Schuster Laboratory, The University,
Manchester M13 9PL, UK
\newline
$^{ 17}$Department of Physics, University of Maryland,
College Park, MD 20742, USA
\newline
% ...
$^{ 23}$Department of Physics and Astronomy, Tel Aviv University,
Tel Aviv 69978, Israel
\newline
$^{ 24}$International Centre for Elementary Particle Physics and
Department of Physics, University of Tokyo, Tokyo 113-0033, and
Kobe University, Kobe 657-8501, Japan
\newline
% ...
$^{  b}$ and Royal Society University Research Fellow
\newline
% ...
\newpage
\section[intro]{Introduction}
The phenomenon of $\BB$ mixing is now well established.
In the case of the $\Bd$ system,
the mass difference, $\dmd$, between the two $\Bd$
mass eigenstates has been 
measured rather precisely~\cite{pdg}.
This mass difference gives the oscillation frequency
between $\Bd$ and $\Bdb$.
Although these measurements can be used to gain
information on the CKM matrix element $V_{\mathrm{td}}$,
this is hampered by large theoretical uncertainties
on both the meson decay constant, $f_{\Bd}$, and
the QCD bag model vacuum insertion parameter, $B_{\Bd}$~\cite{ali}.
This difficulty may be overcome if $\dms$, the $\Bs$ oscillation
frequency, is also measured.
In this case, the CKM information can be extracted
via the relation
\begin{equation}
\frac{\dms}{\dmd} = \frac{m_{\Bs}}{m_{\Bd}} \cdot 
 \frac{|V_{\mathrm{ts}}|^2}{|V_{\mathrm{td}}|^2}
   \cdot \frac{f^2_{\Bs} B_{\Bs}}{f^2_{\Bd} B_{\Bd}} \; ,
\end{equation}
where $m_{\Bs}$ and $m_{\Bd}$ are the $\Bs$ and $\Bd$ masses,
as the ratio of decay constants for $\Bd$ and $\Bs$ mesons is much
better known than the absolute values \cite{ali,nir}.
Information on $|V_{\mathrm{td}}|$ could then be extracted by 
inserting $|V_{\mathrm{ts}}|$, which is relatively 
well known~\cite{pdg}.

The most restrictive of the 
published limits~\cite{mj,ll2,alinc,delphi} 
indicates that $\dms > 9.6$ \ps~\cite{alinc},
while the previous best limit from OPAL~\cite{mj} gives 
$\dms > 3.1$~\ps .

A study of $\Bs$ oscillation is presented in this paper, using
lepton charge to tag the flavour of the $\Bs$ at decay and a 
jet charge technique (Section 3), 
or opposite-jet leptons where available, to tag the $\Bs$
flavour at production.
The analysis studies the oscillation as a function of the
proper decay-time reconstructed using secondary vertices (Section 2), 
using a purity in semileptonic b-decays which is evaluated
event-by-event (Section 4).
Note that the analysis is sensitive to both $\Bs$ and $\Bd$
oscillation, but is optimised for the study of $\Bs$ oscillation;
the parameters describing the $\Bd$ system are input from
previous measurements. 

The analysis technique is similar to that presented in previous 
papers~\cite{mj,ll2}, but includes more data (from 1995), takes
advantage of three-dimensional vertexing, and features a more
sophisticated jet charge definition.
Finally, it combines the single lepton and dilepton data.

\section{Event selection}
The analysis was performed on data collected by the OPAL detector
at LEP in the vicinity of
the $\Z$ peak from 1991 to 1995.
The OPAL detector has been described elsewhere~\cite{OPAL,opalsi}.
Tracking of charged particles is performed by a central detector,
consisting of a silicon microvertex detector,
a vertex chamber, a jet chamber
and $z$-chambers.\footnote{
The coordinate system is defined with
positive $z$ along the $\mathrm{e}^-$
beam direction, $\theta$ and
$\phi$ being the polar and azimuthal angles.
The origin is taken to
be the centre of the detector
and $r$ is the distance to the $z$-axis.}
The central detector is 
positioned inside a
solenoid, which provides a uniform magnetic
field of 0.435 T.
The silicon microvertex detector consists of two layers of
silicon strip detectors;
the inner layer covers a polar angle range of
$| \cos \theta | < 0.83$ and
the outer layer covers $| \cos \theta |< 0.77$.
This detector provided measurements of hits in the $r$-$\phi$ plane
for data taken since 1991, with $z$ coordinates also measured
since 1993. 
The vertex chamber is a precision drift chamber
which covers the range $|\cos \theta | < 0.95$.
The jet chamber is
a large-volume drift chamber, 4.0~m long and 3.7~m in diameter,
providing both tracking and ionisation energy loss (d$E$/d$x$) information.
The $z$-chambers
measure the $z$-coordinate
of tracks as they leave the jet chamber in the range
$|\cos \theta | < 0.72$.
The coil is surrounded by a
time-of-flight counter array and
a barrel lead-glass electromagnetic calorimeter with a presampler.
Including the endcap electromagnetic calorimeters,
the lead-glass blocks
cover the range $| \cos \theta | < 0.98$.
The magnet return yoke 
is instrumented with streamer tubes
and serves as a hadron calorimeter.
Outside the hadron calorimeter are muon chambers, which
cover 93\% of the full solid angle.

Hadronic $\Z$ decays were selected using criteria
described in a previous publication~\cite{TKMH}.
Only data where the silicon microvertex detector was
fully functional 
were accepted, resulting in 2.9 million hadronic $\Z$ decays
with 3-dimensional silicon
readout 
(3D),
and 0.9 million with 
silicon readout 
only in the $r$-$\phi$ projection (2D).
Tracks and electromagnetic clusters 
not associated to tracks were grouped
into jets using
a cone jet algorithm~\cite{conejet}.

The selection of electrons and muons as candidates for semileptonic
decays of b hadrons followed the procedure described
in the previous paper~\cite{mj},
the exception being that 
an improved purity tag at a later stage of the analysis 
allowed
the cut on the lepton neural network $\nn$ to be loosened from
0.7 to 0.5.

Monte Carlo events were generated using the Jetset 7.4 program~\cite{jset}
with parameters tuned to OPAL data~\cite{tune}, and were processed by the
detector simulation program~\cite{gopal}.

\subsection{Proper time reconstruction}
An attempt to reconstruct the proper decay time of the 
parent b hadron was made for each jet containing a lepton
with a minimum $p_t$ of 0.7~\Gc\ relative
to the jet axis.
The algorithm used to reconstruct the decay length of the
supposed parent b-hadron was different from that described 
previously~\cite{mj}.
In principle, each b jet should contain two vertices other than
the primary vertex, corresponding to the decays of the b and c hadrons.
These two vertices are referred to as the secondary and tertiary
vertices, respectively. 
In the new algorithm, the positions of the secondary and tertiary
vertices are allowed to vary independently
in a maximum likelihood fit, 
where the likelihood is calculated
for each vertex position by taking the product of 
likelihoods of all the
tracks in the jet.

Each track contributes a likelihood of 
$(w/2)\times (P_s+P_t) + (1-w)\times P_p$,
where $w$ is the probability for the track to originate
from the secondary vertex, as determined from the track
momentum and angle relative to the jet direction.
For the lepton candidate, $w$ is fixed to 1.
The quantities $P_p$, $P_s$ and $P_t$ are the probabilities for
the track to be compatible with belonging to 
the primary, secondary or tertiary vertices
respectively, based on the impact parameters relative
to the assumed vertex positions both in $r$-$\phi$ and $r$-$z$.

Such a fit has 6 free parameters: the B decay length, the D decay
length and two angles each for the B direction and the D direction.
In practice, the power of the fit to measure the D decay length
was found to be poor. 
The resolution was therefore improved by
imposing a constraint on the D decay length, $L_D$: the likelihood
was multiplied by $\exp (-L_D/L)$ where $L=0.093$~cm was an average
D decay length in the Monte Carlo, and $L_D$ was constrained
to be positive.
The effect of this constraint is 
to put
the fitted D decay length to 0 for most vertices, 
effectively reducing the number of  
free parameters to three.
Gaussian constraints are imposed on the B direction angles, using the
results of a 
B direction-finding algorithm similar to that described in~\cite{robins},
which weighted tracks according to their
rapidity relative to the estimated B direction. 
Loose constraints were also imposed on the D direction angles.
The B decay length was 
constrained\footnote{The constraint was implemented by
subtracting a smooth penalty function
from the log likelihood, which was 0 in the quoted range and parabolic
outside this range.} 
to lie in the range
-0.4~cm to 2.5~cm.

For leptons coming from semileptonic B decays,
according to Monte Carlo with 3D silicon information,
about 4\% of the jets considered were rejected as containing
fewer than 3 tracks passing the quality requirements.
In about 8\% of cases the fit
did not successfully converge, and these candidates
were rejected.
About 2\% of the candidates were rejected with the
fitted B decay length less than -0.4~cm.
To ensure that the results of the fit were stable and unambiguous,
candidates were rejected if the likelihood was improved when the
decay length was increased by 1 standard deviation.
In addition, a scan of the likelihood
was made as a function of the B decay length with the other
parameters kept fixed.
If the log likelihood was within 1.75 of the fitted result for
any B decay length further than two standard deviations from
the central value, the candidate was rejected.
These two requirements rejected about 24\% of the candidates.
Thus, in total, about 62\% of leptons from B decays
were selected with a successful fit.
For these leptons, the B decay length,
$L_{\mathrm B}$, and its uncertainty, $\sigma_L$, were taken directly from
the fit.

Given a reconstructed secondary vertex, the B energy was determined
in a similar way to that described previously~\cite{mj,ll}.
The energy of the jet containing the lepton was reconstructed
using the $\Z$ mass to constrain the event kinematics, and
the estimated contribution from fragmentation particles was
subtracted.
The fragmentation particles were separated from the b-hadron
decay products using momentum, angle and vertex information.
The uncertainty of the B boost, $\sigma_{\beta \gamma}$,
was determined from the estimated uncertainties on the charged 
and neutral fragmentation energies, which were determined
from the estimated probabilities for each track or cluster
to originate from a fragmentation particle.

The selected vertices were split into two classes, A and B,
according to the quality of the proper time reconstruction.
The better quality secondary vertices (class A) were selected 
by requiring that 
\begin{itemize}
\item
the mass of the reconstructed vertex,
using association probabilities for each track as calculated
in the likelihood,
was larger than 0.5~GeV;
\item
the $\chi^2$ per degree of freedom of the vertex configuration
was less than 6;
\item
the angle between the lepton direction and the
vector joining the primary and secondary vertices was larger than
110 mrad;
\item
the ratio of lepton $p_t$ (relative to the jet axis) and momentum $p$
was less then 0.35;
\item
the reconstructed jet energy was required 
to not exceed
the constrained jet energy by more than 2~GeV.
The reconstructed jet energy, computed as the sum of 
energies of all tracks and clusters in
the jet, should normally be smaller than the constrained jet energy
because of the missing neutrino from the semileptonic decay.
\end{itemize}
Class A contains 67\% of the selected vertices for
leptons from B decays in Monte Carlo;
the remaining 33\% constitute class B.

\begin{figure}[tbp]
\centering
\epsfxsize=17cm
\begin{center}
    \leavevmode
    \epsffile[30 390 532 650]{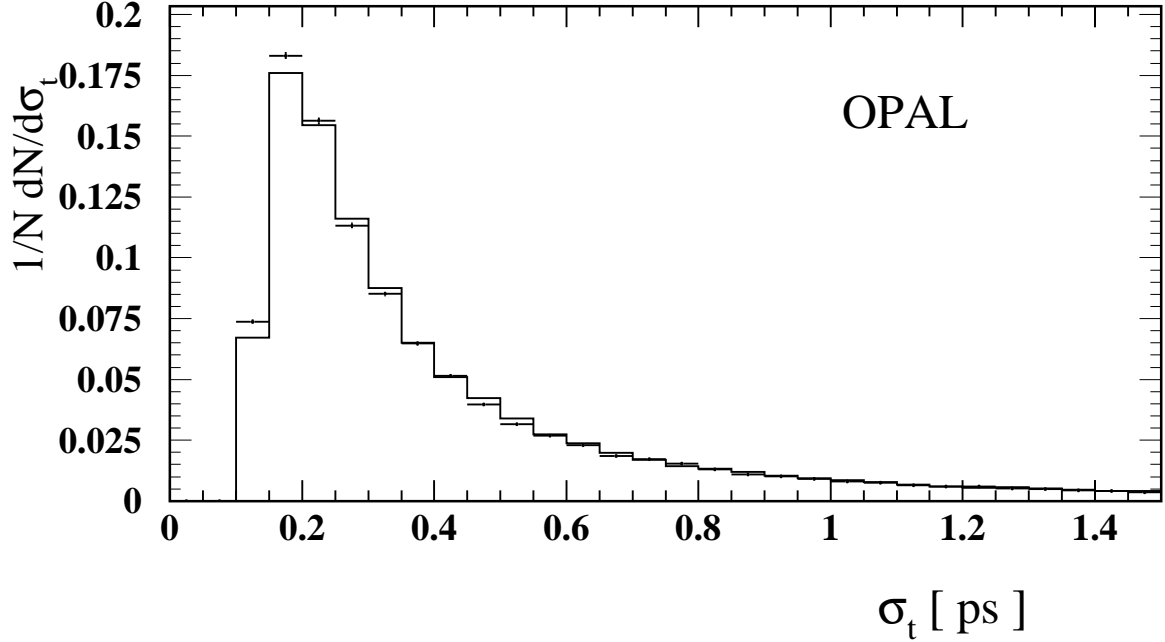}
\end{center}
\vspace{-5 mm}
\caption{ 
The distribution of $\sigma_t$ for all selected vertices 
in the data (points), together with the Monte Carlo prediction
(histogram).
The vertical error bars are obscured by the size of the points.
}
\label{fig:sigt}
\end{figure}

\begin{figure}[htbp]
\centering
\epsfxsize=17cm
\begin{center}
    \leavevmode
    \epsffile[30 400 532 695]{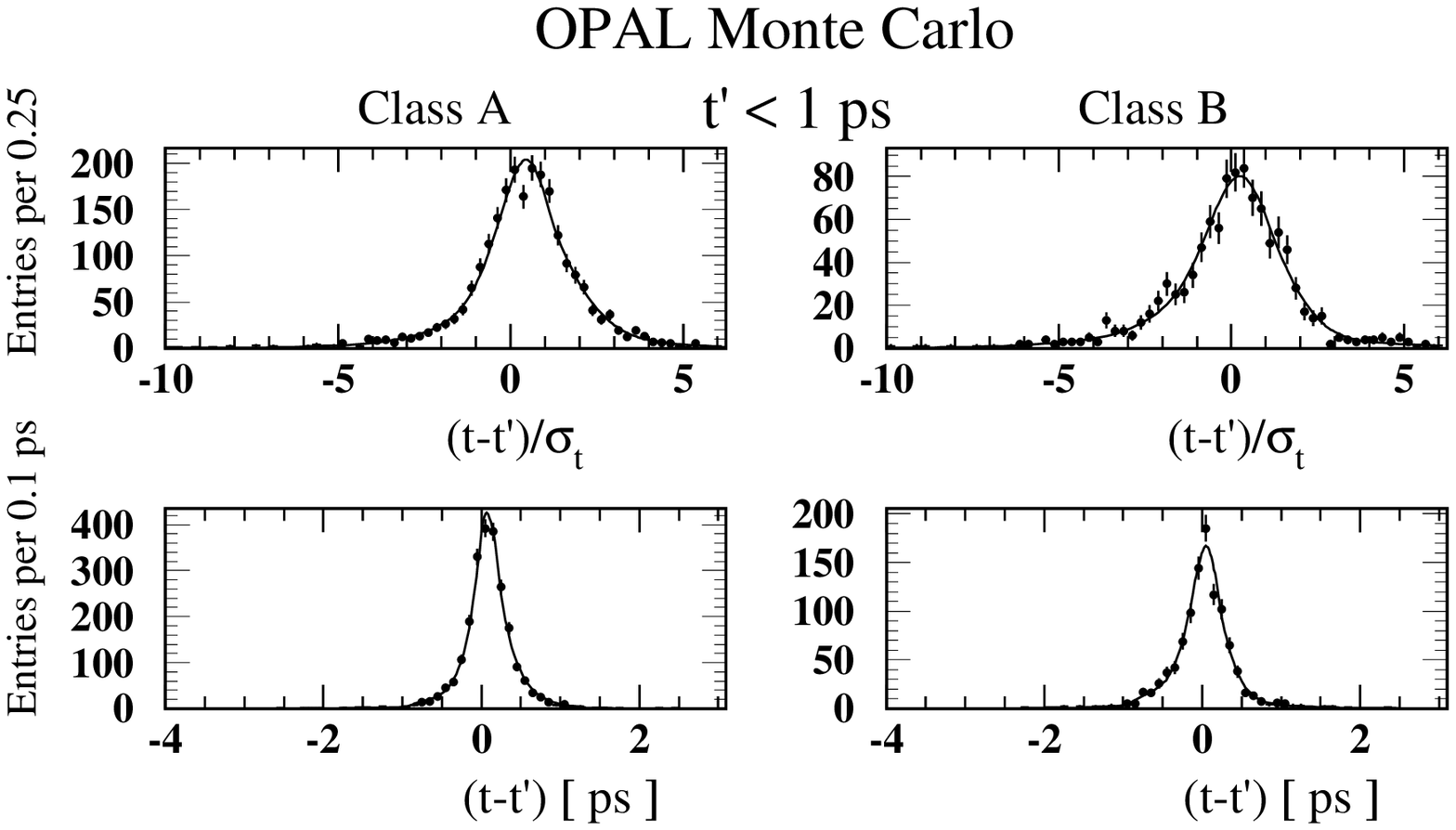}
\end{center}
\epsfxsize=17cm
\begin{center}
    \leavevmode
    \epsffile[30 400 532 675]{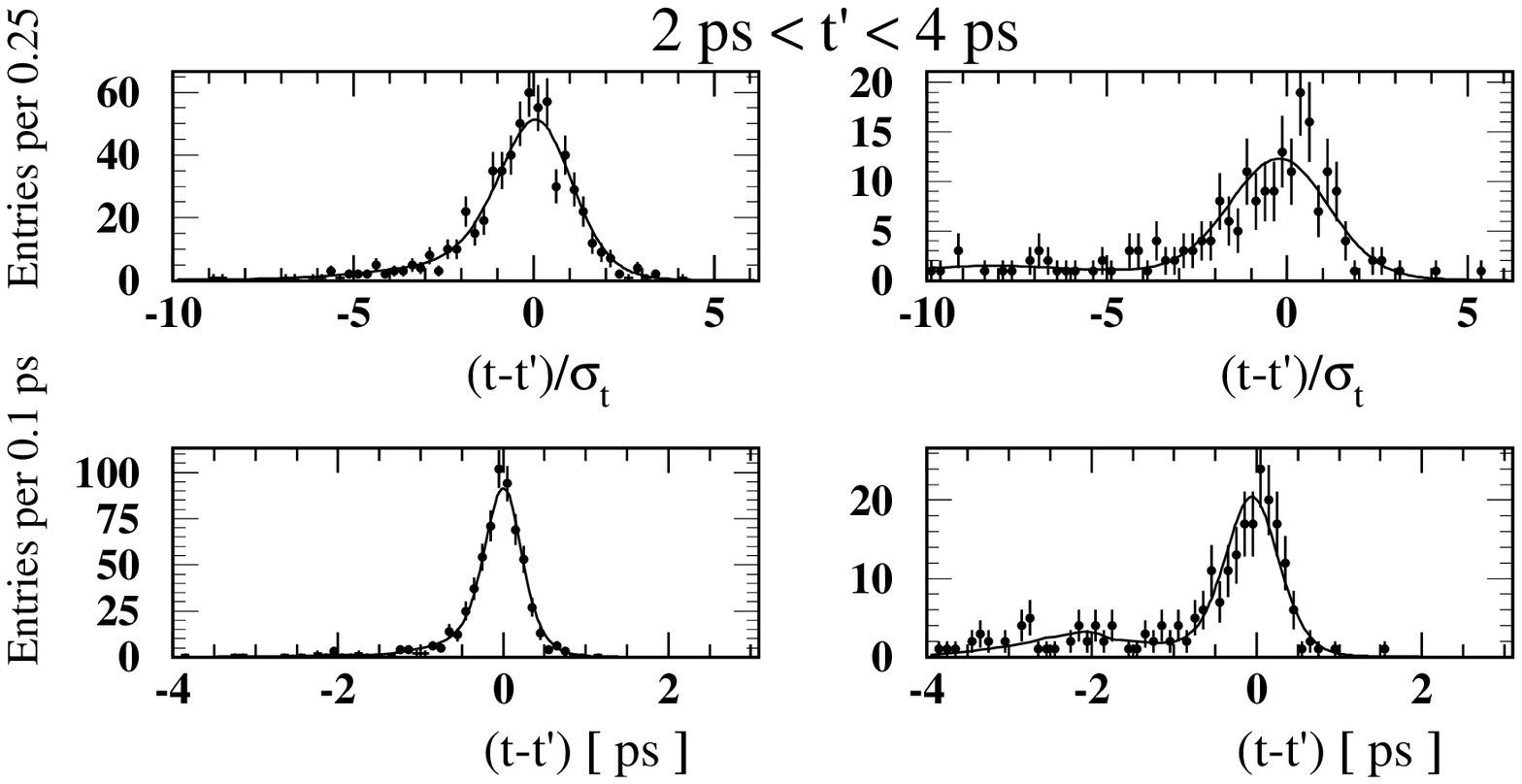}
\end{center}
\vspace{-5 mm}
\caption{ 
The proper time resolution for Monte Carlo $\Bs$ vertices 
with estimated $\sigma_t < 0.3$~ps separated into the
two vertex classes, A and B, for two ranges of true
proper time $t'$.
The curves indicate the fitted resolution functions.
}
\label{fig:tres}
\end{figure}

The proper time, $t$, is given by~\footnote{
We use the notation $\hbar=c=1$.
}:
\begin{equation} t = \frac{L_{\mathrm B}}{\beta \gamma} = 
 \frac{m_{\rm B}}{\sqrt{E_{\rm B}^2 - m_{\rm B}^2}} L_{\mathrm B} \; .\end{equation}
As in the previous analysis~\cite{mj}, use is also made of the 
estimated uncertainty, $\sigma_t$,
on the proper time, calculated from the separately estimated
uncertainties on the decay length, $\sigma_L$, and the boost factor,
$\sigma_{\beta \gamma}$:
 \begin{equation} \left(\frac{\sigma_t}{t}\right)^2 =
   \left(\frac{\sigma_L}{L_{\rm B}}\right)^2 +
   \left(\frac{\sigma_{\beta \gamma}}{\beta \gamma}\right)^2 \; , \end{equation}
where correlations between the uncertainties on $L_{\mathrm B}$ and
$\beta \gamma$ are neglected.
This is unimportant because the shape of the $t$ distribution
is parametrised from Monte Carlo, where such correlations are
included.
The distribution of estimated $\sigma_t$ is shown for the
data for all selected vertices 
in Figure~\ref{fig:sigt}
together with the Monte Carlo prediction.
The slight discrepancy that is visible is not important,
as the analysis uses the value of $\sigma_t$ estimated
for each event.
Plots indicating the proper time resolution for $\Bs$ vertices
with estimated $\sigma_t < 0.3$~ps are shown in Figure~\ref{fig:tres},
together with curves from the fitted resolution functions, defined
as in the previous paper~\cite{mj}.
Note that different resolution functions
are used for class A and class B vertices.
The plots are for the 3D silicon readout; the resolution
functions were determined separately for events with 2D silicon
readout,
where the resolution is similar (though the efficiency is worse).

\begin{figure}[htbp]
\centering
\epsfxsize=17cm
\begin{center}
    \leavevmode
    \epsffile[30 159 532 675]{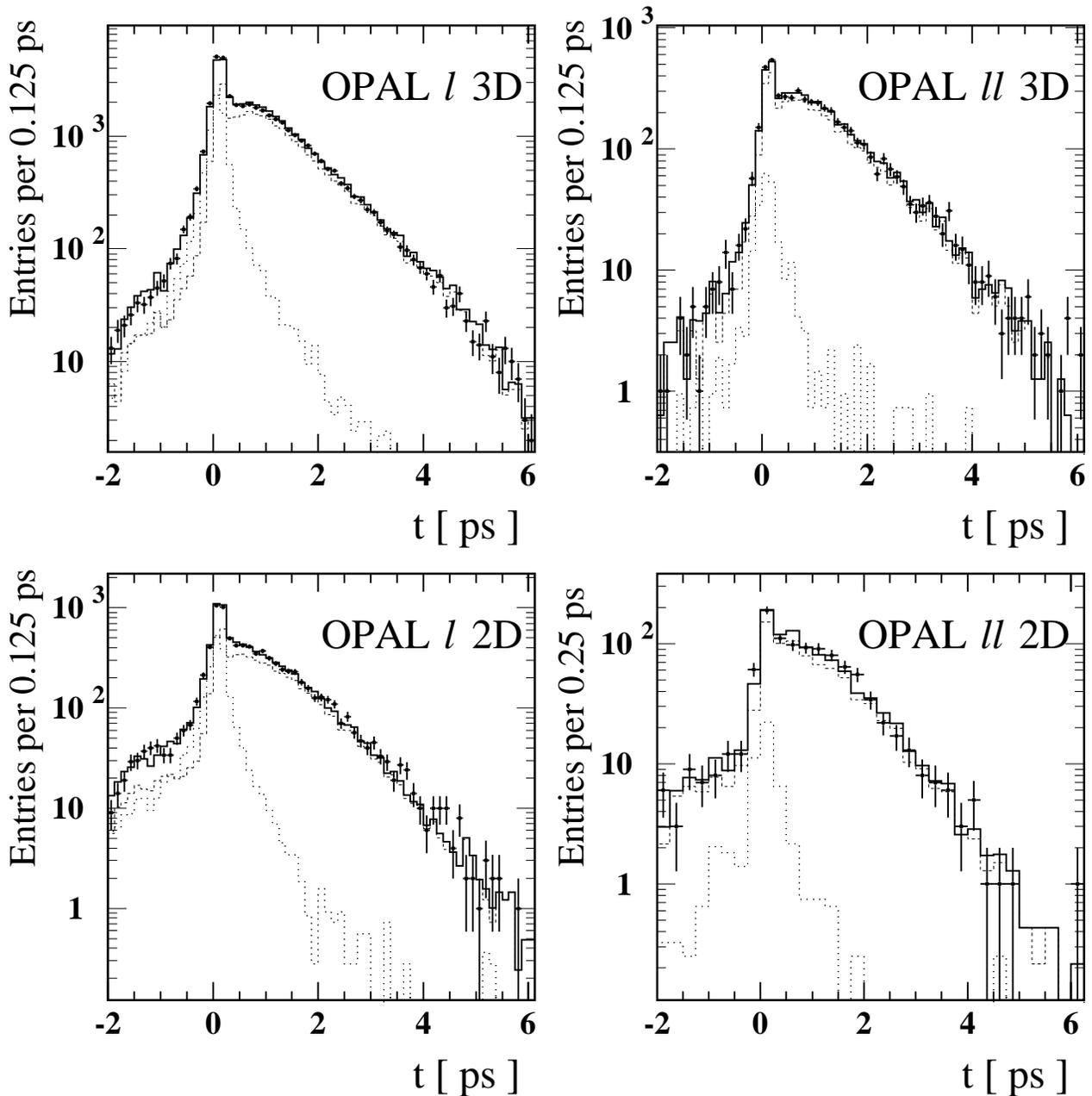}
\end{center}
\vspace{-5 mm}
\caption{ 
The proper time distributions for vertices
with $\sigma_t < 0.3$~ps for single and dilepton
events with 3D and 2D silicon information available (points).
The Monte Carlo predictions are superimposed (solid histograms), together
with the contributions from $\bpr$ decays (dashed) and from
$\uubar$, $\ddbar$, $\ssbar$ or $\ccbar$ events (dotted).
}
\label{fig:trec}
\end{figure}

Events were accepted only if they contained at least one vertex
with $\sigma_t < 0.3$~ps.
This proper time resolution is necessary to 
give sensitivity to $\dms > ~5$~\ps .
A second identified lepton was considered only if separated by
more than $60^\circ$ from the first lepton (the leptons are
ordered using $p$ and $p_t$).
A second vertex in the event,
containing a second identified lepton,
was accepted if it had $\sigma_t < 0.6$~ps.
These requirements led to 43598 events 
selected in the 3D data, of which 5012 had a second identified lepton
and 1788 had two valid vertices.
For the 2D data, 9452 events were selected, of which 1019 had
a second identified lepton and 175 had two valid vertices.
The distributions of reconstructed proper time are shown
in Figure~\ref{fig:trec} for the single lepton and dilepton
events together with the Monte Carlo prediction.
Only one proper time per event,
satisfying $\sigma_t < 0.3$~ps,
is included in the figure for the dilepton
events.
The proper time distributions are given separately for
the 93-95 data and the 91-92 data (only 2D silicon information
available). 
In both cases, the Monte Carlo gives a reasonable description of the data.
Also indicated are the contributions from $\bpr$ decays and from
hadronic $\Z$ decays to
$\uubar$, $\ddbar$, $\ssbar$ or $\ccbar$.
The remaining contributions from $\bbbar$ events, such as $\bcas$
decays and background leptons, are not indicated, but are included
in the total.

\section{Flavour tag}
In order to detect oscillation, one needs to determine the 
b-flavour (b or $\rm \bar{b}$) of the $\Bs$ both at the time
of production and decay.
The flavour at decay is inferred from the charge of the
daughter lepton.
The flavour at production is obtained from the 
charge of a lepton, where available,
in the hemisphere opposite to the first lepton.
For events containing only one identified lepton, it is
determined
by a modified jet charge technique using information in both
hemispheres.
The rest of this section describes the modified jet charge technique.

\subsection{Lepton hemisphere}
The lepton hemisphere is defined by the axis of the jet that contains
the lepton.
For this hemisphere, an unweighted jet charge,
as used in the previous analysis~\cite{mj}, has the desirable property
that the decay products of a neutral B meson contribute a net charge
of zero, provided they are 
fully
contained in the jet.
The jet charge therefore depends only on the net charge of the
fragmentation
tracks, and hence depends on the produced b-flavour.
However, there is extra information that this method does not
utilise.
The leading charged meson from fragmentation should 
reflect the b production
flavour; also, it should have a different
angular distribution and a different momentum spectrum than
the other fragmentation products.
For $\Bs$ production, this leading fragmentation meson
should have an enhanced probability to be a K meson.

To exploit this supplementary information, a number of jet/hemisphere charge
variables were defined:
\begin{itemize}
\item the unweighted jet charge, $\sum_i Q_i$, summing over all
  tracks in the jet;
\item the unweighted hemisphere charge, $\sum_i Q_i$, summing over
  all tracks in the lepton hemisphere;
\item $Q_{\mathrm hem}^{\mathrm NN}$, a weighted hemisphere charge,
 where a neural network is used to assign the weights (see below);
\item $Q_{\mathrm hem}^{\mathrm NN \prime}$, a variant on 
$Q_{\mathrm hem}^{\mathrm NN}$ (see below).
\end{itemize}
Each of these variables was multiplied by the lepton charge, $Q_\ell$,
to provide sensitivity to mixing,
and fed into a neural network, trained to separate
unmixed $\Bs$ decays 
(where the flavour is the same at production and decay)
from mixed $\Bs$ decays, 
with an output $Q_{\mathrm same}$ 
(`same' stands for the same hemisphere as the lepton).
%
%

To determine the weights used for $Q_{\mathrm hem}^{\mathrm net}$, 
a neural network was trained 
on $\Bs$ hemispheres
to distinguish
fragmentation tracks whose charge is opposite to that
of the produced b quark (as expected for the leading charged
fragmention meson)
from
those that have the same charge.
Four inputs were chosen per track: the rapidity relative to the jet,
the track momentum divided by the jet energy and
dE/dx weights for the $\pi$ and K hypotheses. 
B decay products were excluded from the network training.
The neural network output, $\delta$, gives a
weak separation between right and wrong sign fragmentation tracks.
A hemisphere charge was formed to exploit this:
\begin{equation}
Q_{\mathrm hem}^{\mathrm NN} = \sum_i Q_i \times 
(\delta_i -0.5) 
\end{equation}
where the summation is over all tracks in
the same hemisphere as the lepton.
However, the output range with the best separation also includes
a large background
from $\Bs$ decay products.
A second neural network, with output $\epsilon$,
was therefore implemented to separate
fragmentation tracks from B decay products using the track
impact parameters and significances in both $r$-$\phi$ and $r$-$z$ projections.
A variant on the above hemisphere charge was formed: 
\begin{equation}
Q_{\mathrm hem}^{\mathrm NN \prime} = \sum_i Q_i \times 
(\delta_i -0.5) \times \epsilon_i 
\end{equation}
where 
the summation is over all tracks in
the same hemisphere as the lepton.

This hemisphere charge is not optimal, because the important
charge correlations between tracks (especially those
from the $\Bs$ decay) have been neglected.
In fact, when combined with the lepton charge,
the unweighted hemisphere or jet charge gives a 
separation of mixed and unmixed $\Bs$ decays that
is superior to that obtained with $Q_{\mathrm hem}^{\mathrm NN}$
or $Q_{\mathrm hem}^{\mathrm NN \prime}$.
However, the correlation of $Q_{\mathrm hem}^{\mathrm NN (\prime)}$ with
the unweighted hemisphere charge is not very strong, allowing the
combination to achieve a superior performance.

\begin{figure}[htbp]
\centering
\epsfxsize=17cm
\begin{center}
    \leavevmode
    \epsffile[30 150 532 650]{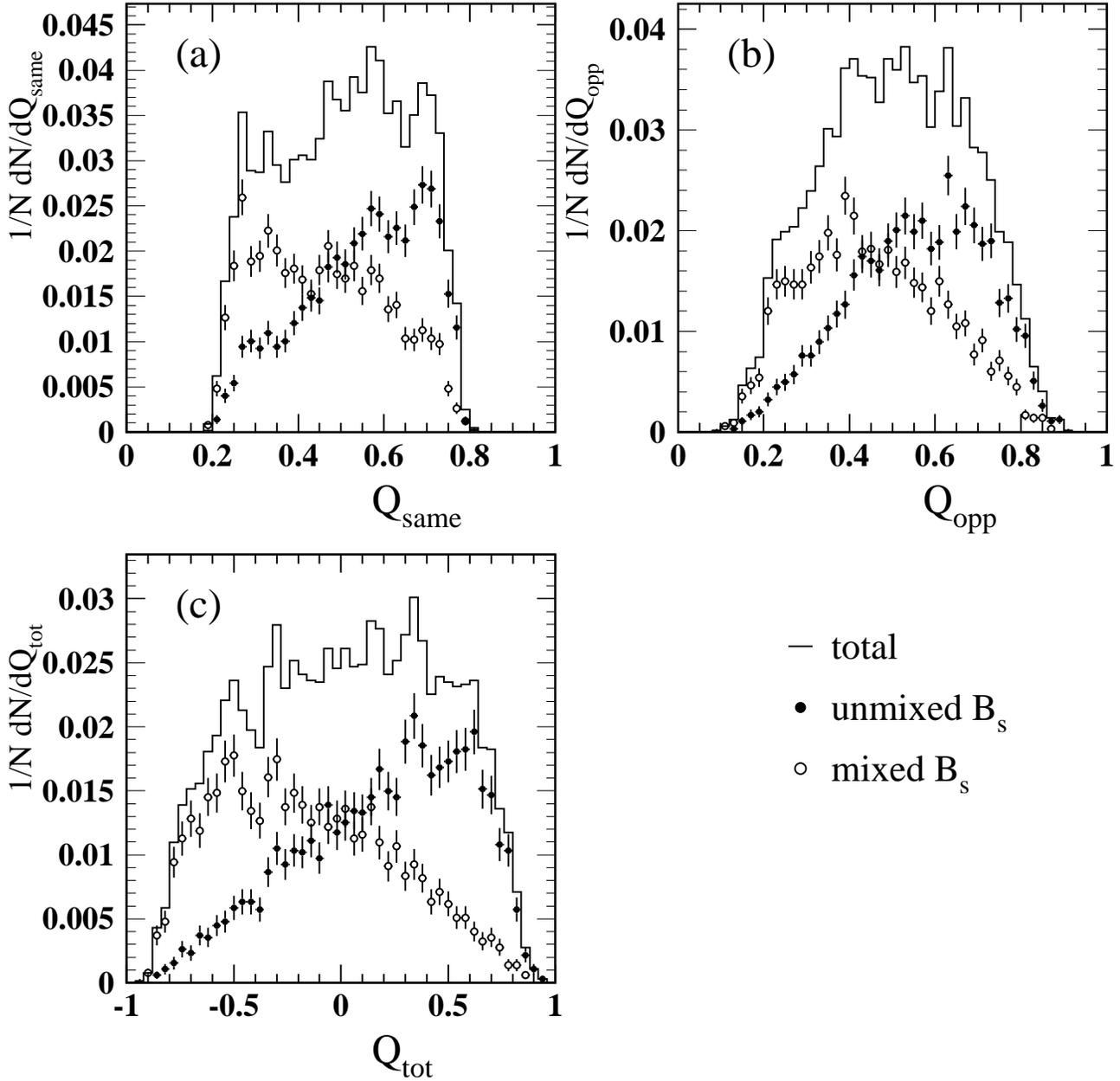}
\end{center}
\vspace{-5 mm}
\caption{ 
The distribution of (a) $Q_{\rm same}$, (b) $Q_{\rm opp}$ and 
(c) $Q_{\rm tot}$
for selected Monte Carlo events where the lepton
originates from a $\Bs$ decay.
The components from unmixed and mixed $\Bs$ decays are indicated.
}
\label{fig:qj3}
\end{figure}

The neural network output for the combination, $Q_{\mathrm same}$, 
is shown in 
Figure~\ref{fig:qj3}(a) for unmixed and mixed $\Bs$ decays.
The power to separate mixed from unmixed $\Bs$ decays
is approximately 40\% better than the unweighted hemisphere 
or the unweighted jet charge alone.
The distribution of $Q_{\mathrm same}$ for the selected data events
is compared to Monte Carlo in Figure~\ref{fig:qtdat}(a), where a 
reasonable agreement is seen.

\subsection{Opposite hemisphere}
The opposite hemisphere is defined by the axis of the
highest energy jet in the event other than the lepton jet.
For this hemisphere, a similar approach was taken,
though the situation is simpler in that 
it is not necessary to distinguish
fragmentation particles from 
B decay products since both
may carry useful charge information.
A number of variables sensitive to the b production flavour were defined:
\begin{itemize}
\item $Q_{\mathrm jet} = \sum_i q_i \left ( \frac{ p_i^l }{E_{\mathrm{beam}}} 
   \right )^\kappa$ where the sum is 
over all tracks in the highest energy jet in 
 the hemisphere using $\kappa=1$ and 
$p_i^l$ is the longitudinal momentum of
   track $i$ relative to the jet axis;
\item $Q_{\mathrm jet}$ using $\kappa=0$;
\item $Q_{\mathrm prob}$ defined below, using a neural network output
for each track in the hemisphere opposite the lepton;
\item vertex charge information - 2 parameters $Q_{\rm vtx}$ and
$\sigma_Q$ (see below).
\end{itemize}
The charge variables were multiplied by the lepton charge and a
neural network,
trained to separate unmixed $\Bs$ decays from mixed $\Bs$ decays,
was used to combine the 5 input variables, giving an output $Q_{\mathrm opp}$.

For the $Q_{\mathrm prob}$ variable,
a neural network was trained to enhance tracks with the desired
charge correlation with the lepton (opposite in the absence of
mixing), using the same four inputs per track as for the track neural
network used in the
lepton hemisphere.
A probability $\varrho$ that the lepton parent was produced as a
$\Bsb$ rather than a $\Bs$ 
is formed for each track, $i$ :
\begin{equation}
\varrho_i = Q_i \times ( \zeta_i - 0.5 ) + 0.5 \; ,
\end{equation} 
where $\zeta_i$ is the output of this neural network for track $i$.
The tracks are combined into an overall probability
per hemisphere, assuming that the probabilities are independent :
\begin{equation}
Q_{\mathrm prob} = \frac{2\times \prod_i \varrho_i}
{\prod_i \varrho_i + \prod_i (1-\varrho_i)} \; 
.
\end{equation}
The power of $Q_{\mathrm prob}$ to separate mixed from unmixed B decays,
when multiplied by the lepton charge, is very similar to that of
the conventional $Q_{\mathrm jet}^{\kappa=1}$
which was used in the previous paper~\cite{mj}.
The two quantities are strongly correlated, but some gain
can still be achieved by combining them.
Excluding the vertex charge information, the neural net combination
of the 3 charge variables  
results in a hemisphere charge with a separation power
23\% better than that of $Q_{\mathrm jet}^{\kappa=1}$ alone.

Further improvement to the opposite hemisphere charge tag
can be made by tagging charged B decays using vertex charge,
as described in a recent paper~\cite{psiks}.
This gave two variables: $Q_{\rm vtx}$ and $\sigma_Q$,
which were available for about 40\% of events
(where vertices significantly separated from the primary vertex
were found).
This information, suitably transformed, was fed into the 5 input
neural network.
The 5 inputs were still used when no vertex information
was available, with the extra 2 inputs set to register
the lack of information.
A further improvement of about 10\% in separation power
was obtained from the
inclusion of the vertex charge information, and the
network output, $Q_{\mathrm opp}$,
is shown in Figure~\ref{fig:qj3}(b) for unmixed
and mixed $\Bs$ decays.
The distribution of $Q_{\mathrm opp}$ for the selected data events,
shown in Figure~\ref{fig:qtdat}(b), is well described
by the Monte Carlo prediction.

\subsection{Combined flavour tag}
\begin{figure}[htbp]
\centering
\epsfxsize=17cm
\begin{center}
    \leavevmode
    \epsffile[30 150 532 650]{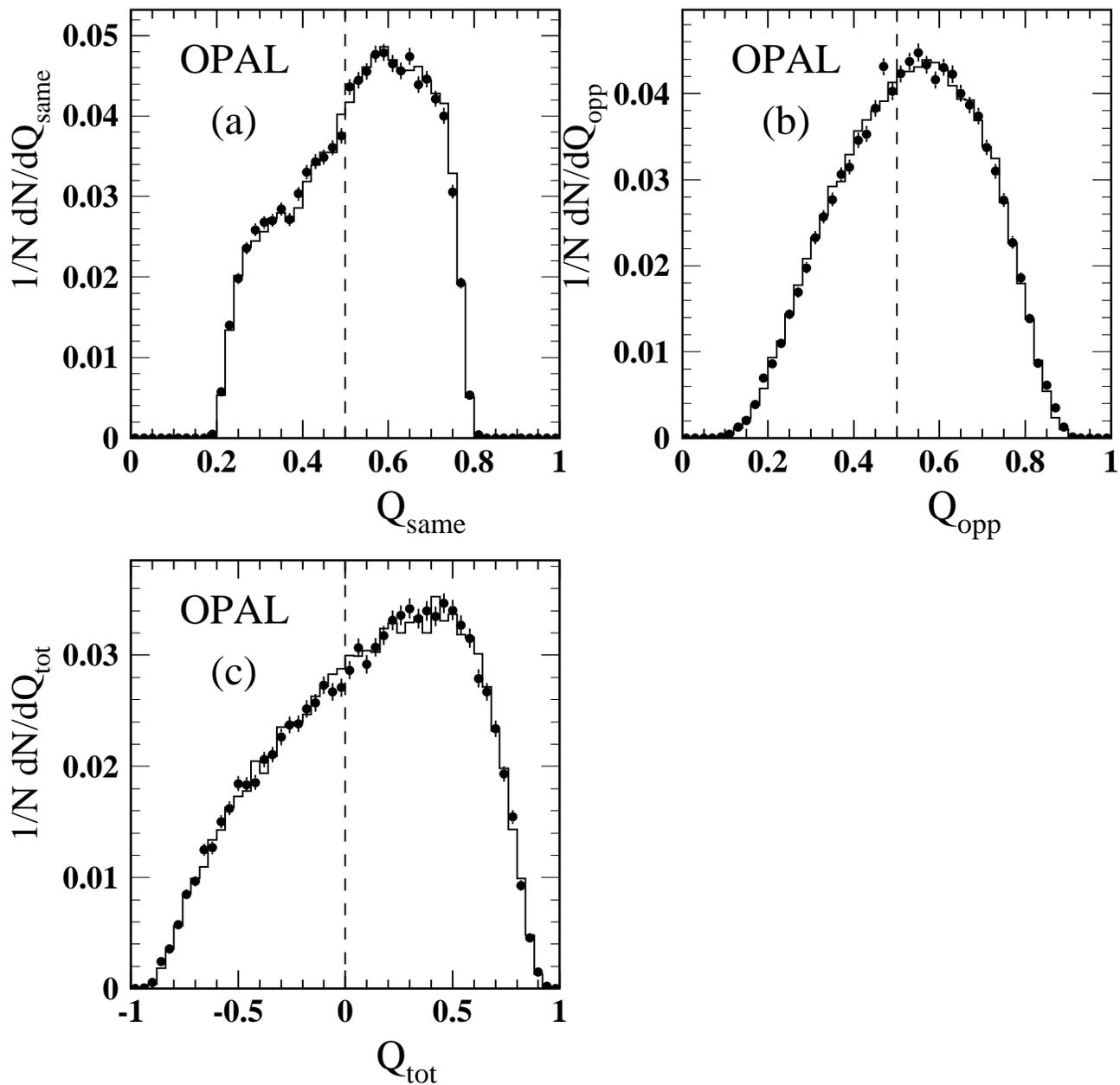}
\end{center}
\vspace{-5 mm}
\caption{ 
The distribution of (a) $Q_{\rm same}$, (b) $Q_{\rm opp}$ and 
(c) $Q_{\rm tot}$ for selected lepton events
in the data (points) together with the Monte Carlo prediction
(histogram).
The dashed line marks the value of the charge variable
where there is no information on the b production flavour. 
}
\label{fig:qtdat}
\end{figure}

The Monte Carlo test samples indicate that the individual values of
both $Q_{\mathrm same}$ and $Q_{\mathrm opp}$ give the 
charge tag purity for $\Bs$ mesons fairly accurately.
Since they are largely independent they may be combined simply to
give a measure of the probability that the $\Bs$ mixed:
\begin{equation}
Q_{\mathrm tot} = \frac{Q_{\mathrm same} \times Q_{\mathrm opp} \times
  2}
{Q_{\mathrm same} \times Q_{\mathrm opp} + 
 (1 - Q_{\mathrm same})(1 - Q_{\mathrm opp})} - 1\; ,
\end{equation}
where the offset of $-1$ is introduced 
so that when $Q_{\mathrm tot} = 0$ 
there is no information on whether the $\Bs$ mixed.
The performance of $Q_{\mathrm tot}$ is shown in
Figure~\ref{fig:qj3}(c),
and has a separation power that is 40\% better than
$Q_{\mathrm 2jet}$, used in the previous analysis~\cite{mj}. 
The distribution of $Q_{\mathrm tot}$ for the selected
data events is shown in Figure~\ref{fig:qtdat}, together
with the Monte Carlo prediction.
The good agreement observed serves as a check on the power
of the charge tag.
%

\section{Tagging the b purity}

The selected lepton candidates contain significant
contributions from sources other than the semileptonic
decays of b hadrons, notably $\bcas$ decays, background
leptons either in $\Zbb$ events or in lighter quark events
and $\mathrm{c \arr \ell}$ decays in $\Zcc$ events~\cite{mj}.
The probability that a lepton candidate originated
from a given source, 
ignoring lifetime information in the lepton jet, 
was evaluated using properties of the lepton jet
and of the opposite hemisphere.
These probabilities are used in the fit (described later)
to scale the likelihood functions corresponding to each
source, where these likelihood functions describe
the reconstructed proper time and charge correlation.

%
For single lepton events,
the value of $\nn$, the lepton neural network output, was used
together with a b-tag from the opposite hemisphere to produce
a single variable that tags the purity of semileptonic decays
of b hadrons.
For dilepton events the purity was determined from the
two values of $\nn$.

\begin{figure}[htbp]
\centering
\epsfxsize=14cm
\begin{center}
    \leavevmode
    \epsffile[30 153 532 675]{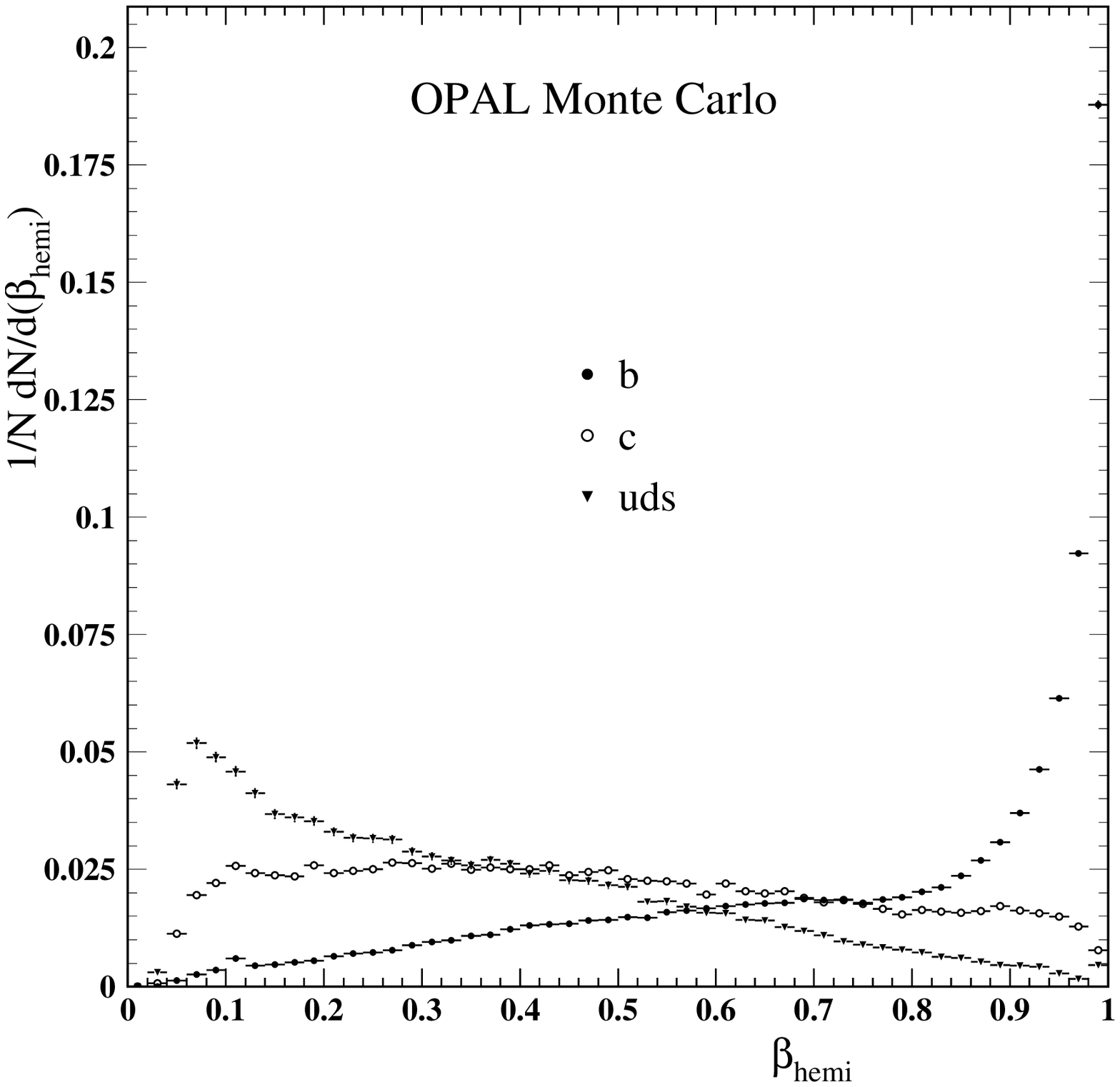}
\end{center}
\vspace{-5 mm}
\caption{ 
Distribution of $\bhemi$ in $\bbbar$ events, $\ccbar$ events and light
quark events.
}
\label{fig:bhemi}
\end{figure}

The opposite hemisphere tag selects the highest energy jet 
in the thrust hemisphere opposite to that containing the lepton jet.
In this jet, 
a neural network was used to separate tracks with 
good quality 3D impact parameter information coming from
b jets from those coming from light quark jets (u, d or s).
%
%
The inputs were
the impact parameters and
their significances in both projections, the distance
of closest approach of the track to the jet axis and the $\chi^2$
for the track to intersect this axis.
%
The neural network outputs for the individual tracks were multiplied together
to form a jet probability, $p_{j \rm 3D}$.
Not all tracks with silicon information in the $r$-$\phi$ plane
have reliable 3D impact parameter information,
so a simpler variable was also constructed, based
solely on the impact parameter significances in the $r$-$\phi$ plane.
In this case, $p_{j \rm 2D}$ was constructed by multiplying the 
likelihoods for each track to be consistent with the primary
vertex. 
The likelihood had a simple double-Gaussian form.
A neural network combined $p_{j \rm 3D}$, $p_{j \rm 2D}$, the numbers
of tracks used for each of these quantities, the $|\cos \theta|$
of the jet axis, and Fox-Wolfram moments which also have
distinguishing power between b jets and light-quark jets.
The neural network was trained to separate b jets from u, d or s jets
(charm events have properties that are intermediate).
The neural network output, $\bhemi$, is shown in Figure~\ref{fig:bhemi}
for $\bbbar$ events and events from lighter quarks.

\begin{figure}[htbp]
\centering
\epsfxsize=17cm
\begin{center}
    \leavevmode
    \epsffile[30 153 532 650]{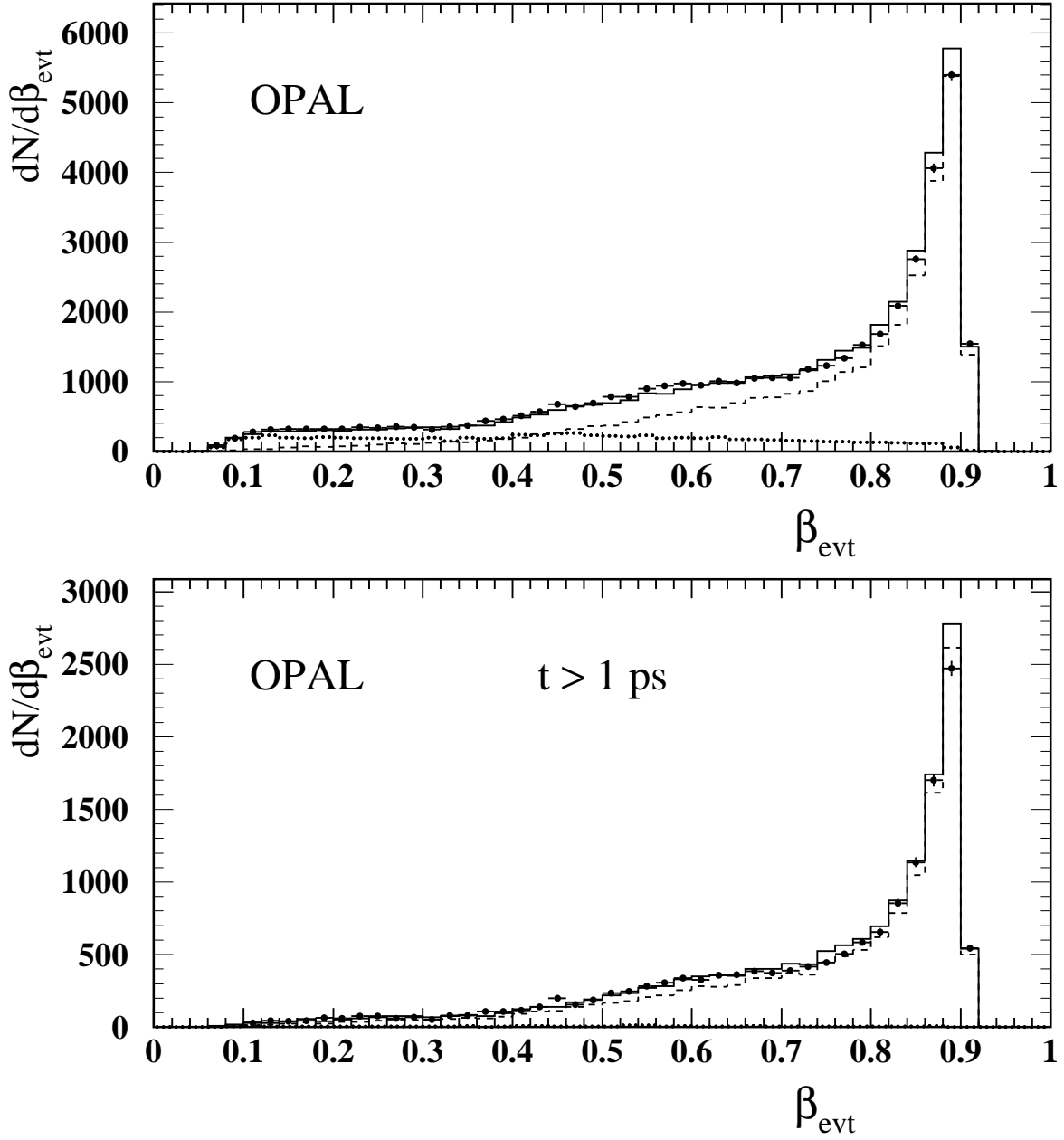}
\end{center}
\vspace{-5 mm}
\caption{ 
The upper plot shows the
distribution of $\btag$ for the data (points) together with
the Monte Carlo prediction (solid line).
The lower plot shows the same thing but with a minimum
reconstructed proper time of 1~ps.
The components from the semileptonic decays
of b hadrons (dashed line) and from $\uubar$, $\ddbar$, $\ssbar$
or $\ccbar$ (dotted line) are
also indicated.
The latter component is almost invisible in the lower plot.
}
\label{fig:btag}
\end{figure}

To form the event b-tag $\btag$, a neural network was used to combine
$\nn$ with $\bhemi$ (though simply multiplying together
the purities obtained
from each tag would yield a similar performance).
The distribution of $\btag$ is shown in Figure~\ref{fig:btag}
for the data, together with the Monte Carlo prediction
and the component from semileptonic decays of b hadrons.
Also shown in the figure are the corresponding $\btag$ distributions
with $t>1$~ps.
A comparison of these two figures indicates that the light quark
background is reasonably modelled.

\section{Fit for \mbox{\boldmath $\dms$}}
A maximum likelihood fit was constructed, similar to
that described previously for single lepton
events~\cite{mj} and for dilepton events~\cite{ll}.
The following quantities were input per event for the
three classes of events :
\begin{enumerate}
\item single lepton events : $t$, $\sigma_t$, $Q_{\rm tot}$ and
  $\beta_{\rm evt}$;
\item dilepton events with one vertex : $t$, $\sigma_t$, 
$Q_\ell^{(1)} \cdot Q_\ell^{(2)}$, $\nn^{(1)}$ and $\nn^{(2)}$;
\item dilepton events with two vertices : 
$t^{(1)}$, $\sigma_t^{(1)}$, $t^{(2)}$, $\sigma_t^{(2)}$, 
$Q_\ell^{(1)} \cdot Q_\ell^{(2)}$, $\nn^{(1)}$ and $\nn^{(2)}$,
\end{enumerate}
where the superscripts are used to differentiate between the two
leptons where appropriate.
As mentioned before, the purities of the different sources
were calculated event-by-event from the appropriate inputs,
taking the distributions of these quantities from Monte Carlo
for each source.

The Monte Carlo predicts that $\bbbar$ events account for 81\%
of the single lepton events, 93.6\% of the dilepton events with
one vertex and 98.3\% of the dilepton events with two vertices.
For the selected vertices in $\bbbar$ events, 87.5\% were
predicted to come from $\bpr$ (or $\bcb$, $\btau$, 
or $\rm b \arr J/\psi\arr \ell$ )~\footnote{
The decays $\bcb$, $\btau$ and half of the 
$\rm b \arr J/\psi\arr \ell^+ \ell^-$
decays
are classed together with $\bpr$ 
because 
they have the same charge correlation between the lepton and the
b quark.}
decays and 9.2\% from $\bcas$ (or $\rm b \arr J/\psi\arr \ell$)
decays.
Of the $\bpr$ decays, an estimated 10.5\% involve $\Bs$ mesons.

In this paper, the only parameter that is varied in the
fit is the $\Bs$ oscillation amplitude, $A$, as defined
in \cite{moser} and used in the previous paper~\cite{mj}.
The probability density for a produced $\Bs$ to decay as a $\Bs$ after time $t^\prime$
is:
\[ {\cal P}_{\rm unmix} = \frac{\exp(-\frac{t^\prime}{\tau})}{\tau}\cdot
   \frac{1 + A\cos \dms t^\prime}{2} 
\]
and to decay as a $\Bsb$ is :
\[ {\cal P}_{\rm mix} = \frac{\exp(-\frac{t^\prime}{\tau})}{\tau}\cdot
   \frac{1 - A\cos \dms t^\prime}{2} \; ,
\]
where $\tau$ is the $\Bs$ lifetime.
At the true oscillation frequency, the fitted value of $A$ should be consistent
with 1.
Far from the true frequency, the expectation value for $A$ is 0.
Therefore, values of $\dms$ may be excluded when $A$ is below 1
and inconsistent with 1.
Since $\dms$ is large, the fitted value of $A$ is relatively
insensitive to sources of systematic uncertainty that affect
the mean $Q_{\rm tot}$ or the overall like-sign fraction in the
case of the dilepton events.
Therefore a complicated fitting scheme, 
as used in the previous paper~\cite{mj},
to reduce the impact of 
these systematic errors is not necessary here.
This also implies that this analysis is not optimal for
$\dms$ values below about 2~\ps .

Unlike the previous analysis~\cite{mj}, the
purity of the b production flavour tag is taken to be independent
of $\dms$ or the fitted amplitude.
The uncertainty on this tagging purity is assessed from the
comparison of data and Monte Carlo shown in Figure~\ref{fig:qtdat}.
This agreement is also sensitive to the overall mixed fraction,
$\chi$, averaged over all b hadrons - which is measured to be 
$0.118 \pm 0.006$~\cite{pdg} 
and the fraction of $\bcas$ decays, taken
to have a 15\% uncertainty~\cite{mj}.

\section{Results}

\begin{figure}[htbp]
\centering
\epsfxsize=17cm
\begin{center}
    \leavevmode
    \epsffile[30 153 532 697]{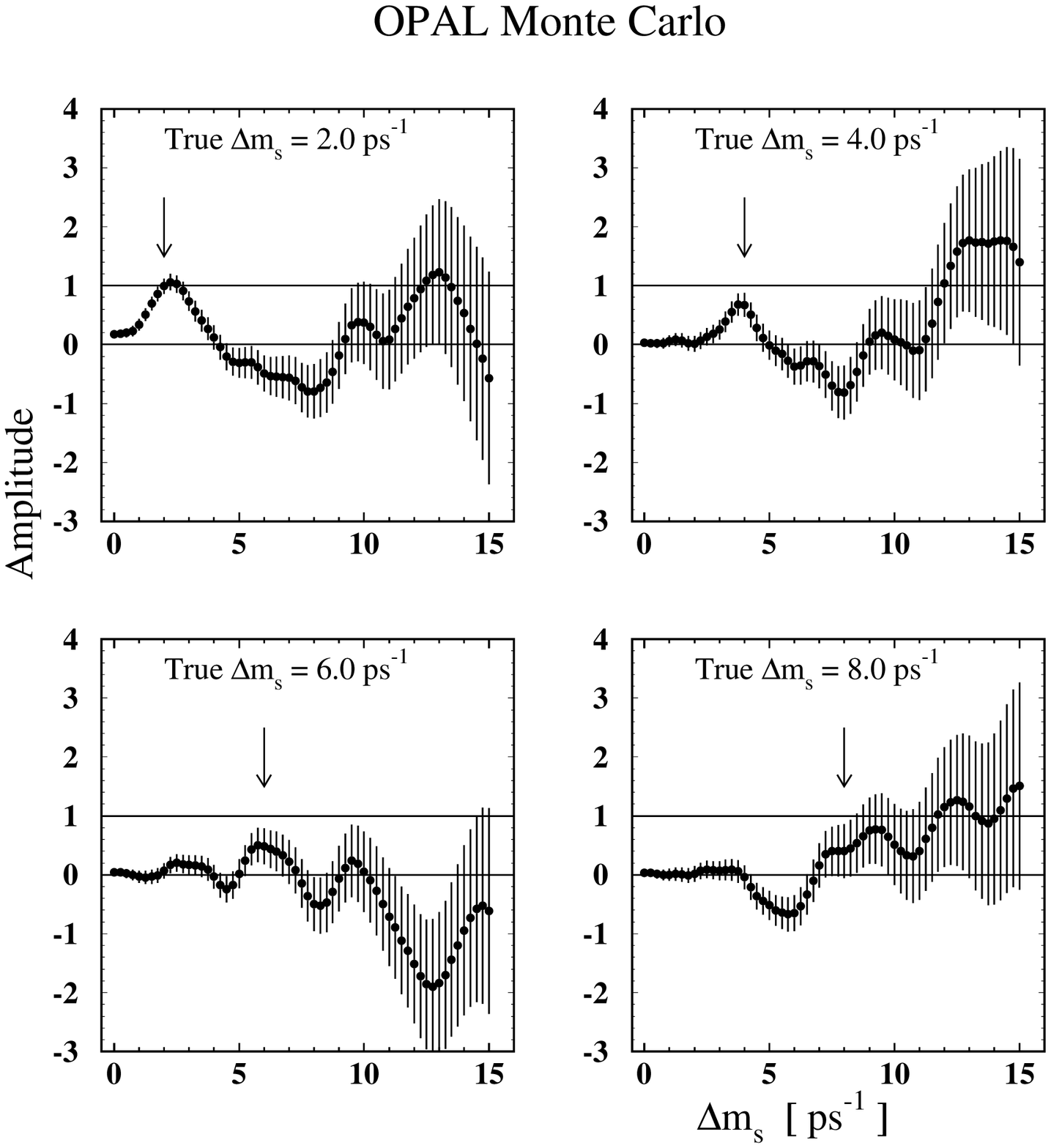}
\end{center}
\vspace{-5 mm}
\caption{ The fitted value of $A$ as a function of $\dms$ for Monte
Carlo data sets simulated using four different $\dms$ values.
The four data sets are statistically correlated.
}
\label{fig:mcscan}
\end{figure}

The performance of the fit was first tested using Monte Carlo
simulated data.
Figure~\ref{fig:mcscan} shows the fitted amplitude
as a function of $\dms$ for four samples generated
with different true $\dms$ values.
These four Monte Carlo samples are statistically correlated,
because the same Monte Carlo events are used to simulate different
oscillation frequencies.
Each represents 6.8 million hadronic Z decays (approximately double
the statistics available in the data). 
The behaviour of these samples is consistent with expectation.
At low values of true $\dms$, a clear peak is seen at 
$A=1$, while at higher frequencies the sensitivity is
insufficient.
In addition, tests were performed for true $\dms$ values
up to 15~\ps\  
using additional Monte Carlo
samples containing only $\Zbb$ events.
The results of these tests, representing greater statistical power,
were also consistent with expectation.

\begin{figure}[htbp]
\centering
\epsfxsize=17cm
\begin{center}
    \leavevmode
    \epsffile[30 153 532 650]{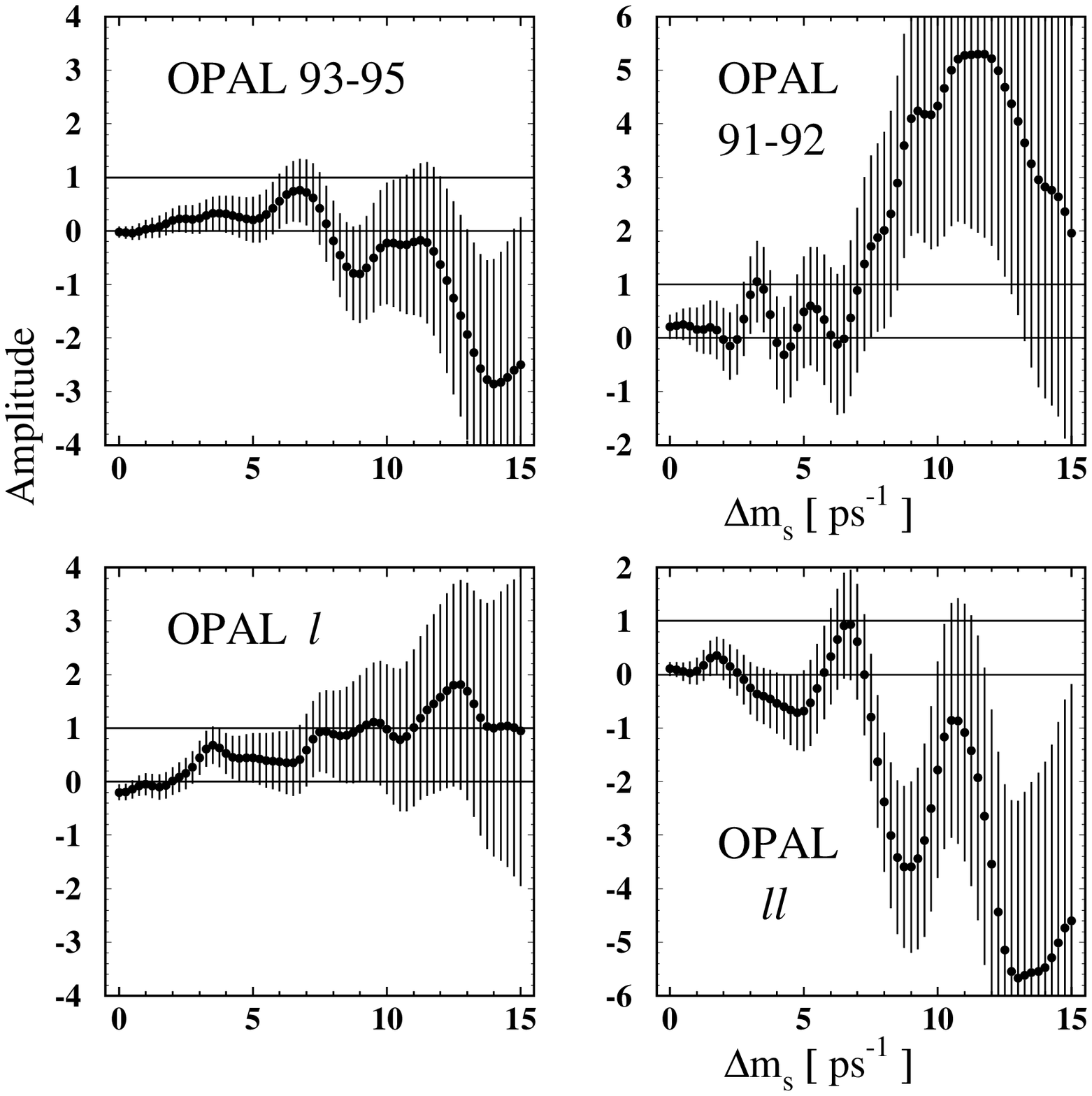}
\end{center}
\vspace{-5 mm}
\caption{ The fitted amplitude as a function of $\dms$ for the data
collected from 93 to 95 (3D silicon) and that collected in 91 and 92
(2D silicon), and for single lepton and dilepton data (where the $91-92$
and $93-95$ data are combined).
The errors are statistical.
}
\label{fig:2d3d}
\end{figure}

For the data, separate results were obtained for the 91-92 data
(2D silicon) (9452 events selected) and the 93-95 data (3D
silicon) (43598 events selected).
These results are shown in Figure~\ref{fig:2d3d}, where the errors
are statistical only.
The data can also be split into single lepton (47109 events)
and dilepton (6031 events) samples.
These results are also shown in Figure~\ref{fig:2d3d}.
It can be seen that the single lepton results are more precise,
except at very low frequency, where the dilepton
events are sensitive to mixing on both sides of the event.
No significant evidence for a signal at any particular
frequency is seen in any of the four plots.
The fact that the measured amplitude is consistent with 0 at
low frequencies, in each case, implies that no systematic
problems are evident.

\subsection{Systematic errors}

Systematic errors arise from a number of sources.
Their effect was determined by varying the appropriate parameter
and obtaining a new set of amplitude results.
This was done using the full data sample, with the exception
of the resolution function uncertainties where Monte Carlo
was also used, as described below. 

The sources of systematic error considered are given in
Table~\ref{tab:sys1}.
\begin{table}[htbp]
\begin{center}
\begin{tabular}{|c|c|} \hline
Parameter & Range \\ \hline
$f_{\mathrm s}$ & $10.5 ^{+1.8}_{-1.7}$\% ~\cite{pdg} \\  
$f_{\rm baryon}$ & $10.1 ^{+3.9}_{-3.1}$\% ~\cite{pdg} \\
$\bcas$ & $\pm 15$\% $\times$ nominal~\cite{mj} \\
$\Zcc$ & $\pm 20$\% $\times$ nominal~\cite{mj} \\
Lepton background & $\pm 20$\% $\times$ nominal~\cite{mj} \\
 & \\
$\langle \tau_{\mathrm b} \rangle $ & $1.564 \pm 0.014$ ps ~\cite{pdg}
\\
$\tau^+ / \taud $ & $ 1.04 \pm 0.04 $ ~\cite{pdg} \\
$\tau_{\mathrm s} / \taud $ & $ 0.99 \pm 0.05 $ ~\cite{pdg} \\
$\tau_{\Lb} / \taud $ & $ 0.79 \pm 0.05 $ ~\cite{pdg} \\
 & \\
$\dmd$ & $0.464 \pm 0.018$~\ps ~\cite{pdg} \\
 & \\
Jet charge & $\pm 0.01$ in tagging purity at $Q_{\rm tot}=0.4$ \\
Resolution function 1 & worsen tracking resolution by 11\% \\
Resolution function 2 & shift $\langle x_E\rangle_{\mathrm b}$ 
by 0.02 \\ \hline
\end{tabular}
\caption{Sources of systematic error.} 
\label{tab:sys1}
\end{center}
\end{table}
%
%
In this table, $f_{\mathrm s}$ and $f_{\rm baryon}$ are 
the production fractions
$f(\rm b \arr \Bs)$ and $f(\rm b \arr$~b~baryon).
The $\bcas$ parameter represents a single scaling factor
for the fraction of events due to $\bcas$ decays.
The $\Zcc$ parameter represents a scale factor for 
the fraction of single lepton events due to $\Zcc$ decays.
The effect is squared for the dilepton events.
The lepton background number is a factor for the 
non-prompt background
rate for each lepton.
The quantity $\tau^+ / \taud $ represents the lifetime ratio for $\Bp$
relative to $\B0$ mesons, and similarly for 
$\tau_{\mathrm s} / \taud$
and $\tau_{\Lb} / \taud $.
These ratios also affect the composition of the sample, as the
semileptonic branching ratios of the individual b hadrons are
taken to be proportional to the lifetimes.
As mentioned in the introduction, the analysis is sensitive
to $\Bd$ as well as $\Bs$ oscillation, so a systematic uncertainty
results from the value of $\dmd$ input.

The systematic error due to the tagging purity of the b production
flavour as determined from jet charge was mentioned above.
The error was modelled by 
changing the relative separation of the mixed and unmixed
distributions of $Q_{\rm tot}$. 
The uncertainty on this offset was assessed using the comparison
of the fraction of data with positive $Q_{\rm tot}$ as a function
of $|Q_{\rm tot}|$
between data and Monte Carlo
simulated events, taking into account uncertainties
in the knowledge of $\chi$ and the fraction of $\bcas$ decays.
The table gives the size of the effect on the tagging purity
at $Q_{\rm tot}=0.4$ as an example.

The resolution functions are determined from 
Monte Carlo (see Figure~\ref{fig:tres}), and
so are affected by uncertainties in the simulation of the data.
Two such effects are considered: firstly, the resolution of
$d_0$ and $\phi_0$ was worsened for all tracks by 11\%,
where $d_0$ is the impact parameter relative to the primary vertex
in the $r$-$\phi$ plane, and $\phi_0$ is the $\phi$ angle at this
point.
This change represents the level of uncertainty in the tracking
resolution.
Secondly, the mean scaled energy of b hadrons, 
$\langle x_E\rangle_{\mathrm b}$, was lowered by 0.02.
Such a change represents a shift of over $2\sigma$ 
with respect to the measured value~\cite{bfrag}, but is inflated
to include the effect of shape uncertainties in the 
b fragmentation function.
Significant statistical fluctuations were observed in the
estimated systematic errors due to the resolution functions.
To reduce this effect, the systematic errors were estimated using
a Monte Carlo sample together
with the data, giving a total sample size 
equivalent to 10.6 million
hadronic $\Z$ decays.
Uncertainty due to charm fragmentation is expected
to have a negligible effect, and
was neglected.

\begin{figure}[htbp]
\centering
\epsfxsize=17cm
\begin{center}
    \leavevmode
    \epsffile[30 153 532 650]{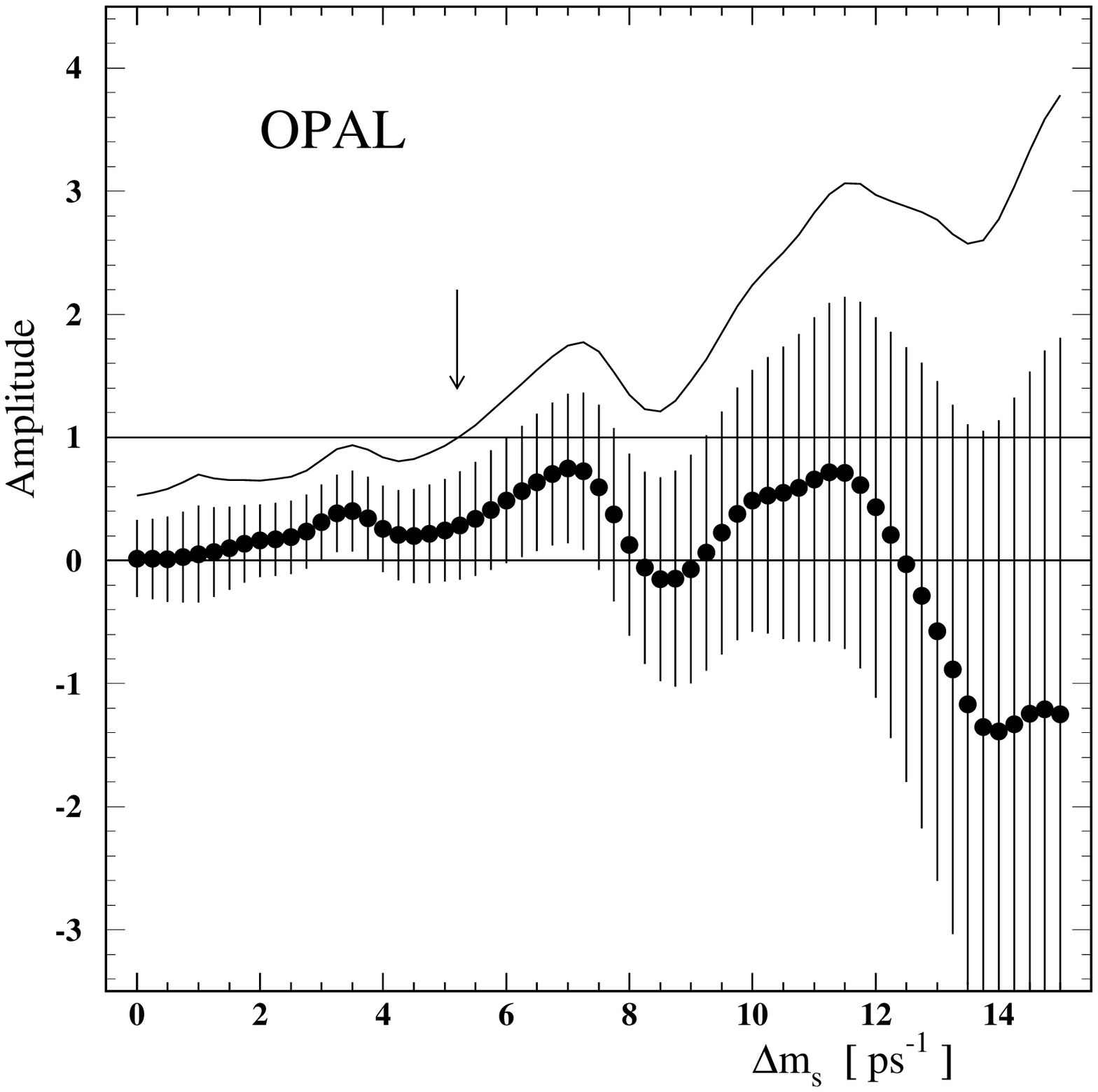}
\end{center}
\vspace{-5 mm}
\caption{ Combined amplitude results for the entire data. 
The errors shown include both statistical and systematic errors.
Also indicated is the `limit' curve which is $A+1.645\sigma_A$.
All regions where this curve lies below $A=1$ are excluded in
method (a).
}
\label{fig:result}
\end{figure}

The individual systematic errors are given for the range of $\dms$
values between 0 and 15~\ps\
in steps of 1~\ps\ in Table~\ref{tab:amp}. 
The sign of the change in amplitude is indicated by $\pm$ or $\mp$
in each case, where $\pm$ indicates that $A$ increases as the parameter
in question increases.
In the case of jet charge, the sign is defined relative to an
increase in the charge tagging purity.
For the resolution function uncertainty due to tracking resolution, 
the sign is defined relative
to the change described above.
The table also includes the overall fitted amplitude and its
statistical error at these points.
From the table it can be seen that the resolution function
uncertainties dominate the systematic error at high frequencies,
while the level of $\bcas$ decays is important at low frequencies.
\begin{table}[htbp]
\begin{center}
{\small
\setlength{\doublerulesep}{0.5cm}
\begin{tabular}{|c|c|c|c|c|c|c|c|c|} \hline
$\dms$ (ps$^{-1}$) & 0 & 1 & 2 & 3 & 4 & 5 & 6 & 7 \\ \hline
$A$ & $  .02$ & $  .05$ & $  .16$ & $  .31$ & $  .26$ & $  .25$ & $
.49$ & $  .75$ \\
$\sigma_A^{\mathrm{stat}}$ & $\pm  .09 $ & $\pm  .16 $ & $\pm  .22 $ &
$\pm  .27 $ & $\pm  .32 $ & $\pm  .39 $ & $\pm  .48 $ & $\pm  .57 $ \\ \hline
$f_{\mathrm s}$ & $\pm  .13 $ & $\pm  .16 $ & $\pm  .07 $ & $\pm  .02
$ & $\pm  .02 $ & $\pm  .01 $ & $\mp  .02 $ & $\mp  .05 $ \\
$f_{\mathrm{baryon}}$ & $\mp  .02 $ & $\mp  .01 $ & $\mp  .01 $ & $\mp  .01 $ & $\mp  .01 $ & $\mp  .01 $ & $\mp  .01 $ & $\mp  .01 $ \\
$\bcas$ & $\pm  .20 $ & $\pm  .24 $ & $\pm  .10 $ & $\pm  .04 $ & $\pm  .02 $ & $\pm  .00 $ & $\mp  .01 $ & $\mp  .01 $ \\
$\Zcc$ & $\mp  .02 $ & $\mp  .04 $ & $\mp  .03 $ & $\mp  .03 $ & $\mp  .03 $ & $\mp  .03 $ & $\mp  .03 $ & $\mp  .02 $ \\
Lepton background & $\pm  .05 $ & $\pm  .09 $ & $\pm  .10 $ & $\pm  .11 $ & $\pm  .11 $ & $\pm  .12 $ & $\pm  .13 $ & $\pm  .15 $ \\
$\langle \tau_{\mathrm b} \rangle $ & $\mp  .00 $ & $\mp  .00 $ & $\mp  .00 $ & $\mp  .00 $ & $\mp  .00 $ & $\mp  .00 $ & $\mp  .00 $ & $\mp  .00 $ \\
$\tau^+ / \taud $ & $\mp  .05 $ & $\mp  .01 $ & $\mp  .01 $ & $\mp  .01 $ & $\mp  .01 $ & $\mp  .01 $ & $\mp  .01 $ & $\mp  .01 $ \\
$\tau_{\mathrm s} / \taud $ & $\pm  .04 $ & $\pm  .03 $ & $\mp  .00 $ & $\mp  .01 $ & $\mp  .01 $ & $\mp  .01 $ & $\mp  .01 $ & $\mp  .01 $ \\
$\tau_{\Lb} / \taud $ & $\mp  .01 $ & $\mp  .00 $ & $\mp  .00 $ & $\mp
.00 $ & $\mp  .00 $ & $\mp  .00 $ & $\mp  .00 $ & $\mp  .00 $ \\
$\dmd$     & $\pm  .05 $ & $\pm  .00 $ & $\pm  .00 $ & $\pm  .01 $ &
 $\pm  .01$ & $\pm  .02 $ & $\pm  .02 $ & $\pm  .02$ \\ 
Jet charge & $\mp  .14 $ & $\mp  .18 $ & $\mp  .10 $ & $\mp  .05 $ & $\mp  .04 $ & $\mp  .03 $ & $\mp  .06 $ & $\mp  .07 $ \\
Tracking resolution & $\pm  .01 $ & $\pm  .02 $ & $\pm  .06 $ & $\pm
.06 $ & $\pm  .02 $ & $\pm  .00 $ & $\mp  .10 $ & $\mp  .07 $ \\
b fragmentation & $\pm  .00 $ & $\pm  .00 $ & $\pm  .05 $ & $\pm  .04 $ & $\pm  .08 $ & $\pm  .05 $ & $\mp  .01 $ & $\mp  .07 $ \\ \hline
$\sigma_A^{\mathrm{syst}}$ &  $\pm  .30 $ & $\pm  .36 $ & $\pm  .20 $ & $\pm  .15 $ & $\pm  .15 $ & $\pm  .14 $ & $\pm  .18 $ & $\pm  .20 $ \\
\hline
\hline
$\dms$ (ps$^{-1}$) & 8 & 9 & 10 & 11 & 12 & 13 & 14 & 15 \\ \hline
$A$ & $  .13$ & $ -.07$ & $  .48$ & $  .66$ & $  .43$ & $ -.57$ & $-1.39$ & $-1.25$ \\
$\sigma_A^{\mathrm{stat}}$ & $\pm  .69 $ & $\pm  .85 $ & $\pm 1.04 $ & $\pm 1.25 $ & $\pm 1.49 $ & $\pm 1.73 $ & $\pm 2.01 $ & $\pm 2.34 $
\\ \hline
$f_{\mathrm s}$ & $\pm  .04 $ & $\pm  .08 $ & $\pm  .03 $ & $\pm  .01 $ & $\pm  .07 $ & $\pm  .25 $ & $\pm  .39 $ & $\pm  .41 $  \\
$f_{\mathrm{baryon}}$ & $\mp  .01 $ & $\mp  .01 $ & $\mp  .01 $ & $\mp  .01 $ & $\mp  .00 $ & $\mp  .00 $ & $\mp  .00 $ & $\mp  .01 $  \\
$\bcas$ & $\mp  .01 $ & $\mp  .01 $ & $\pm  .02 $ & $\pm  .02 $ & $\pm  .02$ & $\pm  .05 $ & $\pm  .07 $ & $\pm  .12 $ \\
$\Zcc$ & $\mp  .03 $ & $\mp  .04 $ & $\mp  .04 $ & $\mp  .04 $ & $\mp  .06 $ & $\mp  .09 $ & $\mp  .11 $ & $\mp  .13 $ \\
Lepton background & $\pm  .15 $ & $\pm  .16 $ & $\pm  .21 $ & $\pm  .25 $ & $\pm  .29 $ & $\pm  .33 $ & $\pm  .37 $ & $\pm  .44 $  \\
$\langle \tau_{\mathrm b} \rangle $ & $\mp  .00 $ & $\mp  .00 $ & $\mp  .00 $ & $\mp  .00 $ & $\mp  .00 $ & $\mp  .00 $ & $\mp  .00 $ & $\pm  .01 $ \\
$\tau^+ / \taud $ & $\mp  .01 $ & $\mp  .02 $ & $\mp  .03 $ & $\mp  .03 $ & $\mp  .03 $ & $\mp  .03 $ & $\mp  .04 $ & $\mp  .06 $ \\
$\tau_{\mathrm s} / \taud $ & $\mp  .00 $ & $\pm  .01 $ & $\pm  .04 $ & $\pm  .03 $ & $\pm  .03 $ & $\pm  .04 $ & $\pm  .06 $ & $\pm  .10 $ \\
$\tau_{\Lb} / \taud $ & $\mp  .00 $ & $\mp  .01 $ & $\mp  .01 $ & $\mp  .01 $ & $\mp  .01 $ & $\mp  .01 $ & $\mp  .01 $ & $\mp  .02 $ \\
$\dmd$     & $\pm  .02 $ & $\pm  .02 $ & $\pm  .02 $ & $\pm  .02 $ &
 $\pm  .02$ & $\pm  .01 $ & $\pm  .01 $ & $\pm  .03$ \\ 
Jet charge & $\mp  .01 $ & $\pm  .02 $ & $\mp  .03 $ & $\mp  .05 $ & $\mp  .02 $ & $\pm  .05 $ & $\pm  .05 $ & $\mp  .01 $ \\
Tracking resolution & $\mp  .09 $ & $\mp  .08 $ & $\mp  .01 $ & $\pm
.05 $ & $\pm  .25 $ & $\pm 0.97 $ & $\pm 1.43 $ & $\pm 1.59 $ \\
b fragmentation & $\mp  .19 $ & $\mp  .31 $ & $\pm  .03 $ & $\pm 0.32 $ & $\pm 0.02 $ & $\mp  .14 $ & $\mp  .09 $ & $\mp 0.95 $
\\ \hline
$\sigma_A^{\mathrm{syst}}$ & $\pm  .26 $ & $\pm  .38 $ & $\pm  .23 $ & $\pm .41 $ & $\pm .40 $ & $\pm 1.07 $ & $\pm 1.54 $ & $\pm 1.96 $ \\
\hline
\end{tabular}
}
\caption{Fit results for
amplitude $A$, with the breakdown of systematic error contributions.} 
\label{tab:amp}
\end{center}
\end{table}
The combined amplitude results for the 91-92 and 93-95 data
together with the total errors are shown in Figure~\ref{fig:result}. 

\subsection{Exclusion regions}
To determine exclusion regions at 95\% confidence level, 
we represent
the measured value of $A$ at a given
value of $\dms$ as a Gaussian distribution
function $G(A-\mu,\sigma_A)$, 
where $\mu$ is the central value and $\sigma_A$ is the
measurement error.
Two alternative methods are then considered to determine 
whether the value of $\dms$ is excluded:
\begin{description}
\item[a)]
values are excluded
where the probability of measuring an amplitude lower than
that observed would be
less than 5\% were that value of $\dms$ the correct
one, i.e. 
\begin{equation}\int_1^{\infty} G(A-\mu,\sigma_A) {\mathrm d}A < 0.05 \; ,
\end{equation} or
\item[b)]
the same definition, but limited to the positive region, i.e.
\begin{equation} 
\frac{\int_1^{\infty} G(A-\mu,\sigma_A) {\mathrm d}A}
        {\int_0^{\infty} G(A-\mu,\sigma_A) {\mathrm d}A} < 0.05 \; .
\end{equation}
\end{description}
The first definition gives 
a true 95\% confidence level, in the sense that there
is a 5\% probability to exclude the true value.
However, it is not protected
against setting limits well beyond the experimental sensitivity.
The second definition makes use of the fact that the predicted value
of $A$ lies between 0 and 1, regardless of the value of $\dms$.
It is automatically protected against setting limits beyond the
sensitive range.
For a true value of $\dms$ well beyond the sensitive range, method (a)
would exclude the true value in 5\% of the experiments, while for method
(b) this percentage would tend towards zero.

For method (a) the excluded regions may be determined simply by
plotting the curve $A+1.645\sigma_A$ as a function of $\dms$
(Figure~\ref{fig:result}).
All regions where the curve lies below $A=1$ may be excluded.
This gives a lower limit $\dms > 5.2$~\ps\ at the 95\% confidence
level.
The limit that would be obtained, were $A$ measured to be 0 at
every value of $\dms$, is 7.0~\ps . 
This is known as the sensitivity of the analysis.

For method (b), the confidence level must be calculated at each
$\dms$ point, according to the above formula.
This gives the result $\dms > 5.0$~\ps\ at the 95\% confidence
level, very similar to the method (a) result.
\section{Conclusion}
Single lepton and dilepton events were used to study
$\Bs$ oscillations with improved sensitivity with respect to
previous OPAL papers~\cite{mj,ll2}.
The experiment was not able to resolve the oscillations, but
can place a lower limit $\dms > 5.2$~\ps\ at the 95\% confidence
level.
This result is consistent with previous 
publications~\cite{mj,ll2,alinc,delphi}, and
supersedes the previously published OPAL
results~\cite{mj,ll2}.
The sensitivity of the analysis (the lower limit that would be expected
were the true oscillation frequency very large) was found to be
7.0~\ps , the second highest relative to previous 
publications~\cite{alinc,delphi}.
%
%
\par
\vspace*{1.cm}
\section*{Acknowledgements}
We particularly wish to thank the SL Division for the efficient operation
of the LEP accelerator at all energies
 and for their continuing close cooperation with
our experimental group.  We thank our colleagues from CEA, DAPNIA/SPP,
CE-Saclay for their efforts over the years on the time-of-flight and trigger
systems which we continue to use.  In addition to the support staff at our own
institutions we are pleased to acknowledge the  \\
Department of Energy, USA, \\
National Science Foundation, USA, \\
Particle Physics and Astronomy Research Council, UK, \\
Natural Sciences and Engineering Research Council, Canada, \\
Israel Science Foundation, administered by the Israel
Academy of Science and Humanities, \\
Minerva Gesellschaft, \\
Benoziyo Center for High Energy Physics,\\
Japanese Ministry of Education, Science and Culture (the
Monbusho) and a grant under the Monbusho International
Science Research Program,\\
Japanese Society for the Promotion of Science (JSPS),\\
German Israeli Bi-national Science Foundation (GIF), \\
Bundesministerium f\"ur Bildung, Wissenschaft,
Forschung und Technologie, Germany, \\
National Research Council of Canada, \\
Research Corporation, USA,\\
Hungarian Foundation for Scientific Research, OTKA T-029328, 
T023793 and OTKA F-023259.\\

\end{document}